\documentclass[12pt]{article}
\begin{document}
\title{Matter waves in terms of the unitary representations of the Lorentz group}
\author{Rudolf A. Frick\thanks{E-mail: rf@thp.uni-koeln.de}\\Institut f\"ur Theoretische Physik,\\Universit\"at zu K\"oln, Z\"ulpicher Str 77, 50937 K\"oln, Germany}
\date{}
\maketitle
\begin{abstract}
 In a generalized Heisenberg/Schroedinger picture, the unitary representations of the Lorentz group may, for a massive relativistic particle, be used to attribute to  waves an extra wavelength that is longer than the de Broglie wavelength. Propagators are defined as spacetime transitions between states with different eigenvalues of the first or the second Casimir operator of the Lorentz algebra.
\end{abstract}


\section{Introduction}
The aim of the present paper is to develop a relativistic formalism for attributing to  matter waves  an extra wavelength that is based on the principal series of the infinite-dimentional unitary irreducible representations (UIR)  of the Lorentz group  and  a  generalized Heisenberg/Schr\"odinger picture. 

The principal series   correspond to the eigenvalues $1+\alpha^2-{\nu}^2,\,(0\leq\alpha<\infty,\quad\nu=-s,...,s)$ of the first Casimir operator $C_1={\bf N}^2-{\bf J}^2$, (${\bf N},\,{\bf J}$ are boost and rotation generators, respectively) and the eigenvalues $\alpha\nu$ of the second Casimir operator $C_2={\bf N}\cdot{\bf J}$ of the Lorentz group. The representations $(\alpha,\nu)$ and $(-\alpha,-\nu)$ are unitarily equivalent  \cite{Wign,Joos,Toll2,Bar}.
  
In \cite{Shap}, it was proposed to classify the states of a relativistic particle by means of  the parameters $\alpha$ and $\nu$ related to  the eigenvalues of the operators  $C_1$ and  $C_2$ and to carry out the  expansion of the  wave function in the momentum space representation over the functions  (${\bf p}$ = momentum, $p_0=\sqrt{m^2c^4+c^2{\bf p}^2}$, $m={\rm mass}$, $\nu=0$, ${\bf n}^{2}({\theta},{\varphi})=1$, $pn=(p_0-{\bf n}c{\bf p}))$
\begin{equation}
\label{1}
{\xi}^{(0)}({\bf p},{\alpha},{\bf n})=(pn/mc^2)^{-1}{e}^{i{\alpha}\ln(pn/mc^2)},
\end{equation}
 which realize the UIR of the  Lorentz group (Shapiro transformation) and are the eigenfunctions of the operator $C_1({\bf p})$. The expansion proposed in \cite{Shap} does not include any dependence on the time $t$ and the space coordinates ${\bf x}$, i.e. it is ``spacetime independent".

  In  \cite{Kad} the expansion over the functions (\ref{1}) was used to  introduce the  ``relativistic configurational'' representation (in the following  ${\rho}{\bf n}$ representation, $\rho={\alpha}{\hbar}/mc$) in the framework  of a two-particle equation of the quasipotential type. In this approach the variable ${\rho}$ was interpreted as the relativistic generalization of a relative coordinate. In \cite{Kad} it was shown that the corresponding operators of the Hamiltonian $H^{(0)}(\rho,{\bf n})$ and the 3-momentum ${\bf P}^{(0)}(\rho,{\bf n})$, defined on the functions $\xi^{\ast(0)}({\bf p},{\rho},{\bf n})$, has a form  of the differential-difference operators (their  explicit form shall be used in what follows, see eqs. (\ref{34}) and (\ref{35})).

In our previous   papers  \cite{Fri1,Fri2,Fri3} it has been shown that the UIR  of the Lorentz group may  be also used in a so-called generalized Heisenberg/Schr\"o\-dinger (GH/GS) picture in which either the analogue of Heisenberg states (wave functions) or the analogue of Schr\"odinger operators of a particle are independent of both the time and the space coordinates t, ${\bf x}$. The  derivatives ${\partial}_{t}$ and $\nabla_{\bf x}$  cannot be presented in these operators.   For the states in the GH picture there must be a spacetime independent expansion. In this context the spacetime independent expansions of the Lorentz group are the expansions of the states in the GH picture. It was shown that if   at first we use the momentum space representation,  then the states and operators of the Poincar\'e algebra can be constructed in the  ${\rho}{\bf n}/{\alpha}{\bf n}$ representation.   The  coordinates $t$, ${\bf x}$   may be introduced in the states or in the operators with the help of the transformation (in the following ${\hbar}=c=1$,  $x_0=t$)
\begin{equation}
\label{2}
 S(x):=\exp[-i({x_0}H-{\bf x}\cdot{\bf P})],
\end{equation} 
where  ${H}$ and ${\bf P}$  are the  Hamilton and momentum operators of the particle  in the ${\rho}{\bf n}/{\alpha}{\bf n}$ or in the ${\bf p}$ representation.
 
Using  the transformation (\ref{2}) (${\bf A}(x)=S^{-1}(x){\bf A}S(x)$, ${\bf A}={\bf N},{\bf J})$ we obtain  the generators of the Lorentz group in the GH picture  
\begin{equation}
\label{3}
{\bf N}(x)={\bf N}+x_0{\bf P}-{\bf x}{H},{\quad}{\bf J}(x)={\bf J}-{\bf x}\times{\bf P}.
\end{equation}
Time and space coordinates  occur equally in these generators.  From this point of view one can see (\ref{3}) as field operators which satisfies 
the  equations 
\begin{equation}
\label{4}
\frac{\partial{N_i(x)}}{\partial{x_0}}=P_i,\quad\frac{\partial{N_i(x)}}{\partial{x_j}}=-H{\delta}_{ij},\quad\frac{\partial{J_i(x)}}{\partial{x_j}}=-{\epsilon}_{ijk}P_k,
\end{equation}
 and the commutation rules of the Lorentz algebra. The  eigenfunctions \\${\xi}({\bf p},{\rho},{\bf n},x)$ of the Casimir operators $C_1({\bf p},x)$ and $C_2({\bf p},x)$  realize the infinite-dimensional UIR of the  field (\ref{3}).

In the present paper we shall show  that the use of the functions ${\xi}({\bf p},{\rho},{\bf n},x)$ and of the  ${\rho}{\bf n}/{\alpha}{\bf n}$ representation makes its possible to attribute to matter waves  an extra wavelength in the one dimensional space $\rho$.   In  section 2  we study the motion of a massive relativistic particle and derive exact propagators with a difference of $\rho$. We will show that for a massive relativistic particle the extra wavelength is longer than the de Broglie wavelength. We also consider a ``wave packet''. We shall show that in one-dimensional motion the difference of $\rho$ may be  connected with the difference of $x_0$ or $x_1$. In section 3  massless particles with spin 0 and spinning particles $(s=1/2,1)$  are considered.  Anticommuting operators which realize boson-boson transformations will be used for the construction of the Hamilton and momentum operators and the  corresponding eigenfunctions for a  massless particle with spin 1.  For a  mass-zero particle we introduce a mass type parameter $\mu$ ($\rho={\alpha}/{\mu}$).  We will show that for a massless particle  two cases are possible for the extra wavelength.

\section{Propagation of a massive particle}

In order to describe the one-dimension propagation of waves with an extra wavelength, we use the functions ($p_0=\sqrt{m^2+{p}^2}$)
\begin{equation}
\label{5}
{\psi}({p},{\rho})={e}^{im{\rho}\ln[(p_0-p)/m]},
\end{equation}
which realize the unitary representation of the one-dimensional Lorentz group   and are the eigenfunctions of the operator $N(p)=ip_0{\partial}_{p}$ ($N$ ${\Longrightarrow}$ $m\rho$).

 The  operators of the Hamiltonian  $H$ and the momentum  $P$ of the particle in the ${\rho}$ representation  defined  on the functions ${\psi}^{\ast}({p},{\rho})$ 
\begin{equation}
\label{6}
H({\rho}){\psi}^{\ast}({p},{\rho})=p_0{\psi}^{\ast}({p},{\rho}),{\quad}P({\rho}){\psi}^{\ast}({p},{\rho})=p{\psi}^{\ast}({p},{\rho}),
\end{equation}
have the form
\begin{equation}
\label{7}
H({\rho})=m\cosh(-\frac{i}{m}{\partial}_{{\rho}}),{\quad}P({\rho})=m\sinh(-\frac{i}{m}{\partial}_{{\rho}}).
\end{equation}

With the help of the operator $S(x)=\exp[-i({x_0}H({\rho})-{x_1}P({\rho}))]$ we can introduce the  states of the particle with  definite momentum in the GS picture in the $\rho$ representation
\begin{equation}
\label{8}
{\psi}^{\ast}(p,{\rho},x)={e}^{-i({x_0}p_0-{x_1}p+m{\rho}\ln[(p_0-p)/m]},
\end{equation}
\begin{equation}
\label{9}
i\frac{\partial}{{\partial}x_0}{\psi}^{{\ast}}(p,{\rho},x)=H({\rho}){\psi}^{{\ast}}(p,{\rho},x),\quad{-i}\frac{\partial}{{\partial}x_1}{\psi}^{{\ast}}(p,{\rho},x)=P({\rho}){\psi}^{{\ast}}(p,{\rho},x).
\end{equation}

Using  (\ref{8}) and the substitutions $(-\infty<\eta{<}\infty)$
\begin{equation}
\label{10}
p_0=m \cosh\eta,{\quad}p=m \sinh\eta,{\quad}\eta=-\ln[(p_0-p)/m],
\end{equation}
we can introduce the  transition amplitude with different values of $\rho$
\begin{eqnarray}
\label{11}
K_{{\rho},{\rho}'}(x,x')&=&\frac{m}{2\pi}\int{\psi}^{\ast}({\eta},{\rho},x)
{\psi}({\eta},{\rho}',x')d{\eta}\nonumber\\&=&\frac{m}{2\pi}
\int{e}^{-im[(x_0-{x_0}')\cosh\eta-(x_1-{x_1}')\sinh\eta-
  (\rho-{\rho}')\eta]}d{\eta}.
\end{eqnarray}
The propagator (\ref{11}) may be rewritten as 
\begin{equation}
\label{12}
K_{{\rho},{\rho}'}(x,x')={e}^{-i[({x_0}-{x_0}')H({\rho})-({x_1}-{x_1}')P({\rho})]}\delta(\rho-{\rho}').
\end{equation}

Consider the  nonrelativistic limit. In the  nonrelativistic limit the operators $H(\rho)-m$ and $P(\rho)$ in (\ref{7})  assume the form
\begin{equation}
\label{13}
H_{nr}=-\frac{1}{2m}\frac{{\partial}^2}{\partial{\rho}^2},\quad{P}_{nr}=-i\frac{\partial}{\partial\rho}.
\end{equation}
 The terms ${\rho}p$ in the phase of the eigenfunctions of  $H_{nr}$ and ${P}_{nr}$ in the GS picture
\begin{equation}
\label{14}
{\psi}_{nr}({\rho},t,x_1)=S(t,x_1){\psi}_{nr}(\rho)={e}^{-i(tp^2/2m-{x_1}p-{\rho}p)},
\end{equation}
allows us to introduce in the one-dimensional ${\rho}$ space an extra wavelength \\${^{(\rho)}{\lambda}}=2{\pi}/p$.

 Numerically ${^{(\rho)}{\lambda}}$ corresponds to the de Broglie wavelength. For the propagator with a difference of $\rho$ we have the expression
\begin{equation}
\label{15}
K_{nr}=\Big[\frac{m}{2{\pi}i(t-t')}\Big]^{1/2}{e}^{\frac{im[(x_1-{x_1}')+(\rho-{\rho}')]^2}{2(t-t')}}.
\end{equation}

In the relativistic case the  terms $m{\rho}\ln[(p_0-p)/m]$ in the phase of the functions ${\psi}^{\ast}(p,{\rho},x)$ leads to an extra wavelength 
\begin{equation}
\label{16}
{^{(\rho)}{\lambda}}=\frac{2\pi}{m{\vert}\ln[(p_0-p)/m]{\vert}},
\end{equation}
or ($p>0$),
\begin{equation}
\label{17}
{^{(\rho)}{\lambda}}=\frac{2\pi}{m\ln[\sqrt{1+(\frac{2\pi}{m{\lambda}_{\rm{dBr}}})^2}+\frac{2\pi}{m{\lambda}_{\rm{dBr}}}]},
\end{equation}
where ${\lambda}_{\rm{dBr}}$ denotes the de Broglie wavelength $({\lambda}_{\rm{dBr}}=2\pi/p)$. It follows from (\ref{17}) that  ${^{(\rho)}{\lambda}}$ is longer than the de Broglie wavelength. 

The integral (\ref{11}) may be expressed in terms of special types of Bessel functions. Note that the analogous integral, but without ${\rho}-{\rho}'$ , appears in the calculation of the commutation and causal functions in the theory of quantized fields and that for ${\triangle}x_0=x_0-{x_0}'$, and ${\triangle}x_1=x_1-x_1'$ different cases should be distinguished. For the case that ${\triangle}x_0>0$,  ${\triangle}x_0>{\triangle}x_1$,  $s=\sqrt{({\triangle}x_0)^2-({\triangle}x_1)^2}$,  the integral may be expressed in terms of the Hankel function ${H}^{(2)}_{\nu}(z)$ 
\begin{equation}
\label{18}
K_{{\rho},{\rho}'}(s,\theta)=-i\frac{m}{2}\exp[-i(\rho-{\rho}')m(\theta-i\pi/2)]H^{(2)}_{-i(\rho-{\rho}')m}(ms),
\end{equation}
where $\theta=\ln[s/({\triangle}x_0+{\triangle}x_1)]$.

Let us now study the properties of the variable $\rho$. For this purpose we consider a state of motion which is characterised by a wave function which represents a  ``wave packet``
\begin{equation}
\label{19}
{\psi}^{\ast}(x_0,x_1,\rho)=\int_{{\eta}_0-\triangle{\eta}}^{{\eta}_0+\triangle{\eta}}A(\eta){e}^{-i[{\omega}(\eta)x_0-{k}(\eta)x_1-m{\rho}{\eta}]}d{\eta},
\end{equation}
instead of the usual wave packet in quantum mechanics
\begin{equation}
\label{20}
{\psi}^{\ast}(x_0,x_1)=\int_{k_0-\triangle{k}}^{k_0+\triangle{k}}A(k){e}^{-i[{\omega}(k)t-{k}x_1]}d{k}.
\end{equation}
 The function (\ref{19}) we can transform as follows 
\begin{equation}
\label{21}
{\psi}^{\ast}(x_0,x_1,\rho)=2A({\eta}_0)\frac{\sin\lbrace[{(\frac{d\omega}{d{\eta}})}_0{{x_0}}-{(\frac{dk}{d{\eta}})}_0{{x_1}}-m\rho]\triangle{\eta}\rbrace}{[{(\frac{d\omega}{d{\eta}})}_0{{x_0}}-{(\frac{dk}{d{\eta}})}_0{{x_1}}-m\rho]}{\psi}^{\ast}({\eta}_0,{\rho},x).
\end{equation}
In (\ref{21}) we replace $x_1$ by  $x_1-{x_1}'$,  $x_0{\rightarrow}x_0-{x_0}'$, and ${\rho}$ by  $\rho-{\rho}'$, respectively. For the case that $x_1-{x_1}'$$=0$, the maximum of the amplitude in front of ${\psi}^{\ast}({\eta}_0,{\rho-{\rho}'},x-x')$ corresponds to the relation
\begin{equation}
\label{22}
\rho-{\rho}'=\frac{1}{m}{\Big(}{\frac{d\omega}{d{\eta}}{\Big)}}_0{(x_0-{x_0}')}. 
\end{equation}
For the case that $x_0-{x_0}'=0$, for the maximum we have
\begin{equation}
\label{23}
x_1-{x_1}'=m({\rho}'-{\rho})/{{\Big(}\frac{dk}{d{\eta}}{\Big)}}_0.
\end{equation}

In quantum mechanics a wave packet like (\ref{20}) is being used  to study the transition from the quantum to the classical theory. The maximum value of  the amplitude in (\ref{21}) corresponds in the classical version  to the relation
\begin{equation}
\label{24}
x_1={x_1}'+\frac{p}{p_0}({x_0}-{x_0}')+\frac{m}{p_0}({\rho}'-\rho).
\end{equation}
We conclude that the propagators with difference of $\rho$  do not have any analogue in the classical version of the motion of a point particle.

In connection with the relations (\ref{22}) to (\ref{24}) we make the following remarks: The functions 
\begin{equation}
\label{25}
{\psi}(p,{\rho},x')={e}^{i({x_0}'p_0-{x_1}'p+m{\rho}'\ln[(p_0-p)/m]},
\end{equation}
in the integrad (\ref{11}), are the eigenfunctions of the boost generator in the GH picture in the momentum space representation $N(x')=S^{-1}(x')ip_0{\partial}_{p}S(x')$, ($N$ ${\Longrightarrow}$ $m{\rho}'$). Note that (\ref{25}) is just a  simple direct product of two functions that realise the representations of two groups: a) the group of the Lorentz transformations, see (\ref{5}), b) the group of translations in 2-dimensional Minkowski space, i.e. the function $\exp[i({x_0}'p_0-{x_1}'p)]$.

In the form 
\begin{equation}
\label{26}
N(x')=ip_0{\partial}_{p}+\frac{1}{2}({x_0}'-{x_1}')(p+p_0)+\frac{1}{2}({x_0}'+{x_1}')(p-p_0),
\end{equation}
the boost generator  can be considered as a solution of string equation
\begin{equation}
\label{27}
({\partial}^{2}_{{{x_0}'^{2}}}-{\partial}^{2}_{{{x_1}'^{2}}})N(x')=0,
\end{equation}
which satisfy the commutation relations of the Poincar\'e algebra in the GH picture
\begin{equation}
\label{28}
[N(x'),p]=ip_0,\quad[p,p_0]=0,\quad[p_0,N(x')]=-ip.
\end{equation}
The transition amplitude (\ref{11}) describes the propagation of waves with the extra wavelength in a field which is presented by the operator $N(x')$. In the context of the relations (\ref{22})-(\ref{24}) and the equation (\ref{27}) one can treat (\ref{11}) and  (\ref{15}) as  propogators which describe the motion of a extended object like a string in terms of the variable $\rho$. 

The concept of a propagator like (\ref{11}) can now be extended to three dimensional problems.  The equations (\ref{4})  may be rewritten in the form   in which the Hamilton and momentum operators of the particle play the role of sources
\begin{equation}
\label{29} 
{\bf \nabla}_{\bf x}\times{\bf J}(x)=\frac{\partial{\bf N}
(x)}{\partial{x_0}}+{\bf P},\quad{\bf \nabla}_{\bf x}\cdot{\bf J}(x)=0,
\end{equation}
\begin{equation}
\label{30} 
{\bf \nabla}_{\bf x}\cdot{\bf N}(x)=-3H,\quad{\bf \nabla}_{\bf x}\times{\bf N}(x)=0.
\end{equation}
From this point of view the generators of the  Lorentz group in the GH picture  represent a field like the gravitational field in  operator form.  The transition amplitude 
\begin{eqnarray}
\label{31}
K(1,2)&=&\frac{1}{(2\pi)^3}\int{\xi}^{\ast}({\bf p},{\rho},{\bf n})S(x-x'){\xi}({\bf p},{\rho}',{\bf n}')\frac{d{\bf p}}{p_0}\nonumber\\&=&\frac{1}{(2\pi)^3}\int{\xi}^{\ast}({\bf p},{\rho},{\bf n},x){\xi}({\bf p},{\rho}',{\bf n}',x')\frac{d{\bf p}}{p_0},
\end{eqnarray}
describes  the propagation of matter waves in this field.

For a particle with spin 0 the functions ${\xi}^{(0)}({\bf p},{\rho}',{\bf n}',x')$ are the eigenfunctions of the operator $C_1({\bf p},x')$, and
\begin{equation}
\label{32}
{\xi}^{{\ast}(0)}({\bf p},{\rho},{\bf n},x)=(pn/m)^{-1}\exp{[-i(p_0x_0-{\bf p}{\bf x}+m{\rho}\ln(pn/m))]}
\end{equation}
are the states of the particle with definite momentum in the GS picture  (${\xi}^{{\ast}(0)}={\xi}^{{\ast}(0)}({\bf p},{\rho},{\bf n},x)$)
\begin{equation}
\label{33}
i\frac{\partial}{{\partial}x_0}{\xi}^{{\ast}(0)}={H}^{(0)}({\rho},{\bf n}){\xi}^{{\ast}(0)};\quad{-i}\frac{\partial}{{\partial}{\bf x}}{\xi}^{{\ast}(0)}= {\bf P}^{(0)}(\rho,{\bf n}){\xi}^{{\ast}(0)},
\end{equation} 
where $H^{(0)}(\rho,{\bf n})$,  ${\bf P}^{(0)}(\rho,{\bf n})$ are the  Hamilton and momentum operators of the particle in the \({\rho}{\bf n}\) representation (${\bf L}:={\bf L}({\bf n})$) \cite{Kad},
\begin{eqnarray}
\label{34}
H^{(0)}(\rho,{\bf n})&=&m\cosh(\frac{i}{m}{\partial}_{\rho})+\frac{i}{{\rho}}{\sinh}(\frac{i}{m}{\partial}_{\rho}) +\frac{{\bf L}^2}{2m{\rho}^2}\exp({\frac{i}{m}{\partial}_{\rho}}),\\
\label{35}
{\bf P}^{(0)}(\rho,{\bf n})&=&{\bf n}\lbrack{H^{(0)}-m\exp({\frac{i}{m}{\partial}_{\rho}}})\rbrack-\frac{{\bf n}{\times}{\bf L}}{\rho}\exp({\frac{i}{m}{\partial}_{\rho}}).
\end{eqnarray}

 The phase of the functions ${\xi}^{{\ast}(0)}({\bf p},{\rho},{\bf n},x)$ yields the expression
\begin{equation}
\label{36}
{^{(\rho)}{\lambda}}=\frac{2\pi}{m{\vert}\ln(pn/m){\vert}}.
\end{equation}
It should be noted especially that  on the basic of the spacetime independent representations of the Lorentz group for the phase in ${\xi}^{{\ast}(0)}({\bf p},{\rho},{\bf n})$ one also can  formally define a  wavelength like (\ref{36}), but in this case ${^{(\rho)}}{\lambda}$ cannot be related to the propagation of the particle and the de Broglie wavelength.

 For the massive spinning particles $(s=1/2, 1)$ the  Hamilton and momentum operators ${H}^{(s)}({\alpha},{\bf n})$,  ${\bf P}^{(s)}({\alpha},{\bf n})$ were constructed in \cite{Fri2,Fri3}. The Hamilton operators satisfy the equations
\begin{equation}
\label{37}
{H}^{(s)}({\alpha},{\bf n})\widetilde{B}^{(s)}({\alpha},{\bf n})=\widetilde{B}^{(s)}({\alpha},{\bf n}){H}^{(0)}({\alpha},{\bf n}),
\end{equation}
where $\widetilde{B}^{(s)}({\alpha},{\bf n})$ are differential-difference operators.  The explizit forms of $\widetilde{B}^{(s)}({\alpha},{\bf n})$ are given in the appendix. The  phase of the  eigenfunctions of  ${H}^{(s)}({\alpha},{\bf n})$,  ${\bf P}^{(s)}({\alpha},{\bf n})$ carry the same terms as the functions ${\xi}^{{\ast}(0)}({\bf p},{\alpha},{\bf n})$.

 For a relativistic particle with spin 0, for the case that  ${\bf n}={\bf n}'$, the transition amplitude (\ref{31})  can also be written in the form  (${x_0}_2-{x_0}_1:=x_0$,  ${\bf x}_2-{\bf x}_1={\bf r}$),
\begin{equation}
\label{38}
K(1,2)=\frac{1}{(2{\pi})^{3}}e^{(-i-m{\rho}')\frac{\partial}{\partial{m\rho}}}\int\frac{d{\bf p}}{p_0}e^{-i(x_0p_0-{\bf r}{\bf p})}{\xi}^{{\ast}(0)}({\bf p},{\rho},{\bf n}).
\end{equation}
In (\ref{38}) we use the expansion \cite{Kad,Ska}
\begin{equation}
\label{39}
{\xi}^{{\ast}(0)}({\bf p},{\rho},{\bf n})=4\pi\sum_{l=0}^{\infty}\sum_{m=-l}^{l}i^{l}{\cal P}_{l}(p_0,\rho)Y_{lm}({\theta}_{\bf n},{\varphi}_{\bf n})Y^{\ast}_{lm}({\theta}_{\bf p},{\varphi}_{\bf p}),
\end{equation}
where ${\cal P}_{l}(p_0,\rho)Y_{lm}({\theta}_{\bf n},{\varphi}_{\bf n})$ are the eigenfunctions of $H^{(0)}(\rho,{\bf n})$. For $l=0$ $(p_0=m\cosh\chi$, $0\leq\chi<\infty)$
\begin{equation}
\label{40}
{\cal P}_{0}(p_0,\rho)=\frac{\sin{m\rho}\chi}{m\rho\sinh\chi}.
\end{equation} 
Substituting (\ref{39}) into (\ref{38}) one obtains a partial expansion in terms of the Legendre polynomials $P_{l}(({\bf r}{\bf n})/{\vert}{\bf r}{\vert})$,
\begin{equation}
\label{41}
K(1,2)=\sum_{l=0}^{\infty}(2l+1)i^{l}K^{(l)}_{{\rho},{\rho}'}(x_0,r)P_{l}(({\bf r}{\bf n})/{\vert}{\bf r}{\vert}),
\end{equation}
where $(\triangle\rho:=m(\rho-{\rho}')-i)$
\begin{equation}
\label{42}
K^{(l)}_{{\rho},{\rho}'}(x_0,r)=\int{e}^{-ix_0p_0}\frac{{i}^l}{2{\pi}^2}j_l(pr){\cal P}_{l}(p_0,\triangle\rho)\frac{p^2}{p_0}dp.
\end{equation}
In (\ref{42}) $j_{l}(pr)$ are the spherical Bessel functions $(p=m\sinh\chi)$. For $l=0$ 
\begin{equation}
\label{43}
K^{(0)}_{{\rho},{\rho}'}(x_0,r)=\frac{1}{2{\pi}^2}\int{e}^{-ix_0p_0}j_0(pr){\cal P}_{0}(p_0,\triangle\rho)\frac{p^2}{p_0}dp,
\end{equation}
where
\begin{equation}
\label{44} 
j_{0}(pr)=\frac{\sin[rm\sinh\chi]}{rm\sinh\chi}.
\end{equation}
For the case that $x_0>0$,  $x_0>r$,  $s=\sqrt{x^2_{0}-r^2}$, $2{\theta}=\ln{[(x_0+r)/(x_0-r)]}$, we obtain   $(b= -m/4{\pi}r\triangle\rho)$, 
\begin{equation}
\label{45}
K^{(0)}_{{\rho},{\rho}'}(x_0,r)=b{e}^{(\pi\triangle\rho/2)}\sin(\theta\triangle\rho)H^{(2)}_{1+im(\rho-{\rho}')}(ms).
\end{equation}

\section{Mass-zero particle}

The spacetime independent expansion of the Lorentz group for a mass-zero particle was introduced in \cite{Zas2}.  In order to describe the  massless particles  with spin 0 and spin 1/2 we use the supersymmertic model in the $\alpha{\bf n}$ representation which was proposed in \cite{Fri3}. In  this model the operator $\widetilde{B}^{(1/2)}(\alpha,{\bf n})$   realize a supersymmetry transformation for the massive particles. In order to construct the supersymmetry generators for the  massless particles, the operator $\widetilde{B}^{(1/2)}$ and other  supersymmetry generators  were separated in two parts   corresponding to the operators with $\exp(\frac{i}{2}{\partial}_{\alpha})$ and operators with $\exp(-\frac{i}{2}{\partial}_{\alpha})$, respectively. This leads to two types of representations of the supersymmetry generators for the massless particles: representations with  $\exp(\frac{i}{2}{\partial}_{\alpha})$ and representations with $\exp(-\frac{i}{2}{\partial}_{\alpha})$. We use the representations corresponding to the operators with $\exp(-\frac{i}{2}{\partial}_{\alpha})$. In this case for the particles with spin 0 and spin 1/2 one can  use the following operators  ($\mu$ is a constant with the dimension of mass)
\begin{equation}
\label{46}
{H}^{-(0)}_{0}=\frac{{\mu}}{2}\Big[\frac{\alpha-i}{\alpha}\Big]e^{-i{\partial}_{\alpha}},\quad{\bf P}^{-(0)}_0={\bf n}{H}^{-(0)}_0,
\end{equation}
\begin{equation}
\label{47}
\widetilde{H}^{-(1/2)}_0=\frac{{\mu}}{2}\Big[\frac{\alpha-\frac{3i}{2}}{(\alpha-\frac{i}{2})}\Big]e^{-i{\partial}_{\alpha}},\quad\widetilde{\bf P}^{-(1/2)}_0={\bf n}\widetilde{H}^{-(1/2)}_0
\end{equation}
as the Hamilton and momentum operators for the particles with spin 0 and spin 1/2, respectively. These  operators  were constructed  with the help of the anticommuting operators 
\begin{eqnarray}
\label{48}
Q^{-}_{1}=\left(\begin{array}{c}\begin{array}{cc}{0}&\widetilde{K}^{-}\\{\widetilde{B}^{-}}&{0}\end{array}\end{array}\right),\quad{Q}^{-}_{2}=\left(\begin{array}{c}\begin{array}{cc}0&i\widetilde{K}^{-}\\-i\widetilde{B}^{-}&0\end{array}\end{array}\right),
\end{eqnarray}
where
\begin{equation}
\label{49}
\widetilde{B}^{-}=\sqrt{{\mu}}D^{\dagger{(1/2)}}({\bf n})e^{-\frac{i}{2}{\partial}_{\alpha}},\quad\widetilde{K}^{-}=\sqrt{{\mu}}\frac{\alpha-i}{\alpha}D^{(1/2)}({\bf n})e^{-\frac{i}{2}{\partial}_{\alpha}},
\end{equation}
\begin{equation}
\label{50}
\widetilde{K}^{-}\widetilde{B}^{-}=2{H}^{-(0)}_{0},\quad\widetilde{B}^{-}\widetilde{K}^{-}=2\widetilde{H}^{-(1/2)}_{0}.
\end{equation}
The matrix $D^{(1/2)}({\bf n})$  contains the eigenfunctions of the operator ${\bf \sigma}{\bf n}/2$ with the eigenvalues $\nu=-1/2,1/2$.

For the eigenfunctions of the operators (\ref{46}) we may choose $(-\infty<\gamma<\infty)$
\begin{equation}
\label{51}
{\Psi}^{-(0)}(\alpha,{\bf n},{\gamma},{\bf n}')=\frac{1}{\sqrt{\pi}{\alpha}}e^{-\gamma+i\alpha\gamma}\delta({\bf n}-{\bf n}').
\end{equation}
The eigenvalues of ${H}^{-(0)}$  are determined by $p_0={\mu}\frac{e^{\gamma}}{2}$, and the eigenvalues of ${\bf P}^{-(0)}$   by  ${\bf p}=p_0{\bf n}'$. 

The operators (\ref{49}) realize the supersymmetry transformations
\begin{equation}
\label{52}
\widetilde{B}^{-}{\Psi}^{-(0)}=\sqrt{2p_0}{\Psi}^{-(1/2)},\quad\widetilde{K}^{-}{\Psi}^{-(1/2)}=\sqrt{2p_0}{\Psi}^{-(0)},
\end{equation}
where
\begin{equation}
\label{53}
{\Psi}^{-(1/2)}(\alpha,{\bf n},\gamma,{\bf n}')=\frac{\alpha}{\alpha-i/2}{D}^{{\dagger}(1/2)}({\bf n}){\Psi}^{-(0)}(\alpha,{\bf n},\gamma,{\bf n}').
\end{equation}
 In the GS picture we have 
\begin{eqnarray}
\label{54}
{\Psi}^{-(0)}(\alpha,{\bf n},\gamma,{\bf n}',x)&=&\frac{1}{{\sqrt\pi}\alpha}{e}^{-\gamma}\delta({\bf n}-{\bf n}')e^{-i({x_0}p_0-{\bf x}{\bf p}-\alpha\gamma)},\\
\label{55}
{\Psi}^{-(1/2)}(\alpha,{\bf n},\gamma,{\bf n}',x)&=&\frac{\alpha}{\alpha-i/2}{D}^{{\dagger}(1/2)}({\bf n}){\Psi}^{-(0)}(\alpha,{\bf n},\gamma,{\bf n}',x).
\end{eqnarray}

Let us now consider  a massless particle with spin 1. As a first step, we use  the part of the  operator  $\widetilde{B}^{(1)}({\alpha},{\bf n})$ which correspond to the operator  with $\exp{(-i{\partial}_{\alpha})}$ [see the appendix]   
\begin{equation}
\label{56}
\widetilde{B}^{-(1)}={\mu}D^{\dagger{(1)}}({\bf n})\Big[1+\frac{i\tau}{\alpha}\Big]e^{-i{\partial}_{\alpha}},\quad(\tau=1-({\bf s}{\bf n})^2),
\end{equation}
and also  introduce the operator
\begin{equation}
\label{57}
\widetilde{K}^{-(1)}={\mu}\frac{\alpha-2i}{{\alpha}^2}(\alpha-i\tau)D^{(1)}({\bf n})e^{-i{\partial}_{\alpha}},
\end{equation}
for which 
\begin{equation}
\label{58}
\widetilde{K}^{-(1)}\widetilde{B}^{-(1)}=4({H}^{-(0)}_{0})^2.
\end{equation}
The matrix $D^{(1)}({\bf n})$ contains the eigenfunctions of the operator  ${\bf s}{\bf n}$, ($\nu=-1,0,1$).

Using the anticommuting operators 
\begin{equation}
\label{59}
{Q}^{-(1)}_{1}=\left(\begin{array}{c}\begin{array}{cc}0&\widetilde{K}^{-(1)}\\\widetilde{B}^{-(1)}&0\end{array}\end{array}\right),\quad{Q}^{-(1)}_{2}=\left(\begin{array}{c}\begin{array}{cc}0&i\widetilde{K}^{-(1)}\\-i\widetilde{B}^{-(1)}&0\end{array}\end{array}\right),
\end{equation}
we find
\begin{equation}
\label{60}
({Q}^{-(1)}_{1})^2=({Q}^{-(1)}_{2})^2=4(H)^2,{\quad}H:=\left(\begin{array}{c}\begin{array}{cc}{H}^{-(0)}_{0}&0\\0&\widetilde{H}^{-(1)}_{0}\end{array}\end{array}\right),
\end{equation}
where the operator $(\widetilde\tau=D^{\dagger(1)}({\bf n}){\tau}D^{(1)}({\bf n}), {\widetilde\tau}^2=\widetilde\tau)$
\begin{equation}
\label{61}
\widetilde{H}^{-(1)}_{0}=\frac{{\mu}}{2}\frac{(\alpha-2i)}{(\alpha-i)}(1+\frac{\widetilde\tau}{{\alpha}^2})e^{-i{\partial}_{\alpha}}
\end{equation}
is defined by the relation
\begin{equation}
\label{62}
\widetilde{B}^{-1}\widetilde{K}^{-1}=4(\widetilde{H}^{-(1)}_{0})^2.
\end{equation}
The operators  $\widetilde{B}^{-1}$, $\widetilde{K}^{-1}$ in (\ref{59}) realize the following boson-boson transformations
\begin{equation}
\label{63}
\widetilde{B}^{-(1)}{\Psi}^{-(0)}={2p_0}{\Psi}^{-(1)},\quad\widetilde{K}^{-(1)}{\Psi}^{-(1)}={2p_0}{\Psi}^{-(0)},
\end{equation}
where 
\begin{equation}
\label{64}
{\Psi}^{-(1)}(\alpha,{\bf n},{\gamma},{\bf n}')={D}^{{\dagger}(1)}({\bf n})\frac {(\alpha+i\tau)}{\alpha-i}{\Psi}^{-(0)}(\alpha,{\bf n},{\gamma},{\bf n}')
\end{equation}
are the eigenfunctions of the operator $\widetilde{H}^{-(1)}_{0}$,
\begin{equation}
\label{65}
\widetilde{H}^{-(1)}_{0}{\Psi}^{-(1)}(\alpha,{\bf n},{\gamma},{\bf n}')=p_0{\Psi}^{-(1)}(\alpha,{\bf n},{\gamma},{\bf n}').
\end{equation}
For this reason $\widetilde{H}^{-(1)}_{0}$ can be identified with the Hamilton operator of a massless particle with spin 1. 

In order to conctruct other generators, one can use the operators of the Lorentz algebra $({\bf J}^{(0)}={\bf L}$,  $\widetilde{\bf J}^{(1)}=D^{{\dagger}(1)}({\bf n})({\bf L}+{\bf s})D^{(1)}({\bf n}))$
\begin{equation}
\label{66}
{\bf N}^{(0)}:={\bf n}(\alpha-i)+{\bf n}\times{\bf J}^{(0)},{\quad}\widetilde{\bf N}^{(1)}:={\bf n}(\alpha-i)+{\bf n}\times\widetilde{\bf J}^{(1)}
\end{equation}
in the form of
\begin{equation}
\label{67}
{\bf J}:=\left(\begin{array}{c}\begin{array}{cc}{\bf J}^{(0)}&0\\0&{\widetilde{\bf J}^{(1)}}\end{array}\end{array}\right), \quad{\bf N}:=\left(\begin{array}{c}\begin{array}{cc}{\bf N}^{(0)}&0\\0&{\widetilde{\bf N}^{(1)}}\end{array}\end{array}\right).
\end{equation}
For the commutators $[{Q}^{-(1)}_{1},{\bf N}]$ and $[{Q}^{-(1)}_{2},{\bf N}]$ we have
\begin{equation}
\label{68}
[{Q}^{-(1)}_{1},{\bf N}]:={\bf G}_{1},{\quad}[{Q}^{-(1)}_{2},{\bf N}]:={\bf G}_{2},
\end{equation}
and we find $(r=1,2)$
\begin{equation}
\label{69}
\lbrace{Q}^{-(1)}_{r},{\bf G}_{r}\rbrace=-8i{H}\left(\begin{array}{c}\begin{array}{cc}{\bf P}^{-(0)}_0&0\\0&\widetilde{\bf P}^{-(1)}_{0}\end{array}\end{array}\right),
\end{equation}
where $\widetilde{\bf P}^{-(1)}_{0}={\bf n}\widetilde{H}^{-(1)}_{0}$ can be identified with the  momentum operator of the particle with spin 1,
\begin{equation}
\label{70}
\widetilde{\bf P}^{-(1)}_{0}{\Psi}^{-(1)}(\alpha,{\bf n},{\gamma},{\bf n}')=p_0{\bf n}'{\Psi}^{-(1)}(\alpha,{\bf n},{\gamma},{\bf n}').
\end{equation}
For the functions ${\Psi}^{-(1)}(\alpha,{\bf n},{\gamma},{\bf n}',x)=S(x){\Psi}^{-(1)}(\alpha,{\bf n},{\gamma},{\bf n}')$ we have
\begin{equation}
\label{71}
{\Psi}^{-(1)}(\alpha,{\bf n},{\gamma},{\bf n}',x)={D}^{{\dagger}(1)}({\bf n})\frac {(\alpha+i\tau)}{\alpha-i}{\Psi}^{-(0)}(\alpha,{\bf n},{\gamma},{\bf n}',x).
\end{equation}
Using the phase of the functions (\ref{54}), (\ref{55}) and (\ref{71}), we can finally write the extra wavelength in the ${\rho}$ space (${\rho}={\alpha}/{\mu}$)
\begin{equation}
\label{72}
{^{(\rho)}{\lambda}}=\frac{2\pi}{{{\mu}}{\vert}\gamma{\vert}},\quad{^{(\rho)}{\lambda}}=\frac{2\pi}{{\mu}{\vert}\ln(2{\vert}{\bf p}{\vert}/{\mu}){\vert}},
\end{equation}
or $({\vert}{\bf p}{\vert}=2\pi/{\lambda})$
\begin{equation}
\label{73}
{^{(\rho)}{\lambda}}=\frac{2\pi}{{\mu}{\vert}\ln[4\pi/({\mu}{\lambda})]{\vert}}.
\end{equation}
An important point is that for a  massless particle of this wavelength two cases are posibble. For the redshift $z=({^{(\rho)}{\lambda}}-{\lambda})/{{\lambda}}$ we have
\begin{equation}
\label{74}
z=\frac{2\pi}{{\mu}{\lambda}{\vert}\ln[4\pi/({\mu}{\lambda})]{\vert}}-1.
\end{equation}
Numerically, this expression yields that below of about ${\lambda}=2.8430{\times}(2\pi/{\mu})$ we have ${^{(\rho)}{\lambda}}>\lambda$, accordingly $z>0$. For ${\lambda}>2.8430{\times}(2\pi/{\mu})$ one has ${^{(\rho)}{\lambda}}<\lambda$ and $z<0$.

In a ``wave packet'' like (\ref{21}), and in  the relations (\ref{22})-(\ref{24}) we must replace $\eta$ by $\gamma$ and $m$ by $\mu$.

\section{Conclusion}
We have shown that in  a generalized Heisenberg/Schr\"odinger  picture the principal series of the unitary representation of the Lorentz group may be used to describe the propagation of matter waves with an extra wavelength. We found that for a  massive relativistic particle this wavelength is longer than the de Broglie wavelength.   We have shown that  waves with the extra wavelength  may be used to describe the motion of  extended objects like strings. We found transition amplitudes which describe the propagation of waves with the extra wavelength. For the  mass-zero particles with spin 0 and spin 1  we have introduced  anticommuting operators which realize boson-boson transformations. In this way we found new Hamilton and momentum operators  and corresponding eigenfunctions  for a massless particle with spin 1.  We found that for a massless particle  two cases are possible for the extra wavelegth. A specific feature of this wavelength  is that  in this approach the redshift does not correspond to a Doppler effect. We hope that the formalism developed here will be convenient for solving problems in particle physics and astrophysics.

\section{Appendix} 
The operators $\widetilde{B}^{(s)}({\alpha},{\bf n})$ in (\ref{37}) can be  written as follows: for s=1/2 $({\bf L}:={\bf L}({\bf n}))$
\begin{equation}
\label{A.1}
\widetilde{B}^{(1/2)}({\alpha},{\bf n})=\sqrt{m}D^{\dagger(1/2)}({\bf n})\Big[(1-\frac{i\sigma{\bf L}}{\alpha-i/2})e^{\frac{i}{2}{\partial}_{\alpha}}+e^{-\frac{i}{2}{\partial}_{\alpha}}\Big],
\end{equation}
and for s=1 $(\tau=1-({\bf s}{\bf n})^2)$
\begin{eqnarray}
\label{A.2}
\widetilde{B}^{(1)}({\alpha},{\bf n})&=&{m}D^{\dagger(1)}({\bf n})\Big\lbrace[1-\frac{i\tau}{\alpha}-\frac{1+i\alpha+\tau}{{\alpha}(\alpha-i)}2{\bf s}{\bf L}+\frac{i\tau+\alpha}{{\alpha}^2(\alpha-i)}{\bf L}^2\nonumber\\&&-\frac{2({\bf s}{\bf L})^2}{\alpha(\alpha-i)}]e^{i{\partial}_{\alpha}}+{2-\frac{2i{\bf s}{\bf L}}{\alpha-i}+[1+\frac{i\tau}{\alpha}]e^{-i{\partial}_{\alpha}}\Big\rbrace}.
\end{eqnarray}
The matrix $D^{(1/2)}({\bf n})$  contains the eigenfunctions of the operator ${\bf \sigma}{\bf n}/2$ ($\nu=-1/2,1/2$), and the matrix $D^{(1)}({\bf n})$ the eigenfunctions of  ${\bf s}{\bf n}$ ($\nu=-1,0,1$), respectively.

\end{document}